\documentclass[12pt]{article}
\usepackage{amsmath,amssymb,dsfont}
\newcommand{\beq}{\begin{equation}}
\newcommand{\eeq}{\end{equation}}
\newcommand{\ba}{\begin{array}}
\newcommand{\ea}{\end{array}}
\newcommand{\bea}{\begin{eqnarray}}
\newcommand{\eea}{\end{eqnarray}}
\newcommand{\bean}{\begin{eqnarray*}}
\newcommand{\eean}{\end{eqnarray*}}

\newtheorem{theorem}{Theorem}[section]
\newtheorem{prop}[theorem]{Proposition}

\newtheorem{proof}{Proof.}

\newcounter{appendix}
\newcommand{\CH}{{\cal H}}

\newcommand{\CW}{{\cal W}}
\newcommand{\CZ}{{\cal Z}}

\newcommand{\CG}{{\cal G}}
\newcommand{\CM}{{ \cal M}}

\newcommand{\CL}{{\cal L}}

\newcommand{\CY}{{\cal Y}}

\def\be{\beta}
\def\al{\alpha}

\def\l{\lambda}
\def\la{\lambda}

\def\sig{\sigma}

\newcommand{\half}{\frac{1}{2}}

\newcommand{\til}{\tilde}

\def\mat2#1#2#3#4{{\left(\begin{array}{cc}#1 & #2\\ #3 & #4
      \end{array}\right)}}

\def\mats2#1#2#3#4{{\left(\begin{array}{cc}#1 & #2\vspace{2truemm} \\ #3 & #4
\end{array}\right)}}

\def\endpf{\begin{flushright}$\square$\end{flushright}}

\begin{document}
\begin{center}
{\huge B\"acklund transformations for the \\
\vskip0.2truecm
rational Lagrange chain}
\end{center}
\vspace{0.3truecm}
\begin{center}
{\large
Fabio Musso$^\diamondsuit$, 
Matteo Petrera$^\sharp$, Orlando Ragnisco$^\flat$, Giovanni Satta$^\natural$}
\vskip0.3truecm
Dipartimento di Fisica E. Amaldi \\
Universit\`a degli Studi di Roma Tre\\
and\\
Istituto Nazionale di Fisica Nucleare,
Sezione di Roma Tre\\
Via della Vasca Navale 84, 00146 Roma, Italy\\
\vspace{0.3truecm}
E--mail$^\diamondsuit$:  musso@fis.uniroma3.it\\
E--mail$^\sharp$:  petrera@fis.uniroma3.it\\
E--mail$^\flat$:  ragnisco@fis.uniroma3.it\\
E--mail$^\natural$:  satta@fis.uniroma3.it\\
\end{center}
\vspace{0.2truecm}
\abstract{\noindent
We consider a long--range homogeneous chain
where the local variables are the generators of the direct sum
of $N$ $\mathfrak{e}(3)$ interacting Lagrange tops. We call this classical integrable model
rational ``Lagrange chain'' showing how one can obtain it starting from
$\mathfrak{su}(2)$ rational Gaudin models.
Moreover we  construct one- and two--point integrable maps (B\"acklund transformations).
}\vskip 1truecm \noindent

\section{Introduction}

In \cite{MPR} we have performed a In\"on\"u--Wigner contraction on Gaudin models \cite{G1,G2},
showing that the integrability property is preserved
by this algebraic procedure. Starting from rational, trigonometric and elliptic 
Gaudin models it is possible to obtain
new integrable long--range chains, associated to
the same linear $r$-matrix structure. 

The first natural extension \cite{MPR,KPR} of the $N$-sites $\mathfrak{su}(2)$ Gaudin model is obtained
contracting two copies of the Lie-Poisson algebra $\mathfrak{su}(2)$, namely
\beq
\mathfrak{su}(2) \oplus \mathfrak{su}(2) \simeq \mathfrak{o}(4) \rightarrow \mathfrak{e}(3)
\nonumber
\eeq
where $\mathfrak{e}(3)$ is the real euclidean Lie--Poisson algebra in the 3-space. In this way 
one obtains new integrable chains where local variables are the generators of the direct sum
of $N$ $\mathfrak{e}(3)$ interacting Lagrange tops. From this feature comes 
the name ``Lagrange chains''.

Other interesting structures can be inherited from the Gaudin models 
as well. For example in \cite{KPR} it is shown how to obtain B\"acklund
transformations for the Lagrange top from those of the $\mathfrak{su}(2)$ Gaudin model. 
It turns out that it
is possible to generalize this approach to the construction of B\"acklund
transformations for Lagrange chains. This is the main goal of the present paper.

\section{Algebraic contractions of $\mathfrak{su}(2)$ Gaudin models }

In this section we illustrate the contraction procedure performed on the rational
$\mathfrak{su}(2)$ Gaudin model. 
We stress the fact that this construction is also
possible when considering trigonometric and elliptic solutions of the classical Yang--Baxter
equation and considering a generic finite--dimensional simple Lie algebra instead of $\mathfrak{su}(2)$.
See \cite{MPR} for details.

The rational $\mathfrak{su}(2)$ Gaudin model is derived from the following $2 \times 2$ Lax matrix
\beq
L_\CG(\la)=\tau +  \sum_{i=1}^N \sum_{\al=1}^3
\sigma^\al \, \frac{x^\al_i}{(\la-\la_i)} , \label{jj}
\eeq
where $\tau$ can be chosen as any $\mathfrak{su}(2)$-matrix and 
$\sigma^1,\sigma^2,\sigma^3$ as a basis of the 
fundamental representation of $\mathfrak{su}(2)$:
$$
\sigma^1 \doteq  \frac{{\rm{i}} \, \sigma_x}{2} = \half \left(\begin{array}{cc}
0 & {\rm{i}}  \\
{\rm{i}} & 0
\end{array}\right), \quad
\sigma^2 \doteq  \frac{{\rm{i}} \, \sigma_y}{2} = \half \left(\begin{array}{cc}
0 & 1  \\
-1 & 0
\end{array}\right), \quad
\sigma^3 \doteq \frac{{\rm{i}} \, \sigma_z }{2}= \half \left(\begin{array}{cc}
{\rm{i}} & 0  \\
0 & -{\rm{i}}
\end{array}\right),
$$
where $\sigma_x,\sigma_y,\sigma_z$ are the Pauli matrices.
The constants $\l_i \in \mathbb{C}$ are parameters
of the model and 
$\l \in \mathbb{C}$ is the spectral parameter.
The local variables of the model $x^\al_i $, $\al=1,2,3$, $i=1,...,N$ are the real generators 
of the direct sum of $N$ $\mathfrak{su}(2)$ spins with the
following Lie--Poisson brackets:
\beq
\left\{ x^\al_i, x^\be_j \right\} = \delta_{ij} \, x^\gamma_i \qquad i,j=1,...,N \nonumber
\eeq
where $\al \be \gamma$ is the cyclic permutation of 123.

The Lax matrix (\ref{jj}) satisfies the linear $r$-matrix Poisson algebra:
$$
\left\{L_\CG(\l) \otimes \mathds{1}, \mathds{1} \otimes L_\CG(\mu) \right\} -
\left[r(\l - \mu), L_\CG(\l) \otimes \mathds{1} + \mathds{1} \otimes L_\CG(\mu) \right]=0,
\label{333}
$$
where $\mathds{1}$ is the $2 \times 2$ identity matrix and the $r$-matrix is given by
$$
r(\la)=\frac{{\rm{i}}}{ \l} \sum_{\al=1}^3  \sig^{\al} \otimes \sig^{\al}.
\label{666}
$$

Performing a In\"on\"u--Wigner contraction to the direct sum of 2 copies of $\mathfrak{su}(2)$
we obtain from (\ref{jj}) the following new Lax matrix \cite{MPR}:
\beq
L (\l) = \tau  + \sum_{\alpha=1}^3
\sig^{\alpha} \left[\frac{y^{\al}}{\la} + \frac{z^{\al}}{\la^2}
\right], \label{mat}
\eeq
where the new real generators $(y^{\al}, z^{\al})$, $\al=1,2,3$ 
satisfy the $\mathfrak{e}(3)$ Lie--Poisson brackets:
\beq
\left\{    y^{\alpha},y^{\beta}   \right\}= y^{\gamma}, \qquad
\left\{    y^{\alpha},z^{\beta}   \right\}=  z^{\gamma}, \qquad
\left\{    z^{\al},z^{\be}   \right\}= 0 , \qquad \alpha,\beta,\gamma= 1,2,3, \label{e3} \nonumber
\eeq
where $\al \be \gamma$ is the cyclic permutation of 123.
Note that in the rational case the Lax matrix (\ref{mat}) is exactly
the Lax matrix of the Lagrange top \cite{Au,RSTS}. This particular contraction procedure
on the two--site $\mathfrak{su}(2)$ rational Gaudin model has been considered in \cite{KPR}.

We now extend the Lax matrix (\ref{mat}) to the 
$N$-bodies case. Namely we consider the 
Lax matrix associated to the Lie--Poisson algebra of the direct sum of $N$ copies of
$\mathfrak{e}(3)$:
\beq
\CL(\l)= \tau +  \sum_{i=1}^N   \sum_{\alpha=1}^3
\sig^{\alpha} \left[\frac{y_i^{\al}}{\l - \l_i}
+  \frac{z_i^{\al}}{(\l - \l_i)^2 } \right],\label{lax} \nonumber 
\eeq
where the local variables of the model are generators of the direct sum of $N$ $\mathfrak{e}(3)$ tops,
$(y_i^{\al}, z_i^{\al})$, $i=1,...,N$, $\al=1,2,3$ 
with the following Lie--Poisson brackets:
\beq
\left\{    y_i^{\alpha},y_j^{\beta}   \right\}= \delta_{ij} \, y_j^{\gamma}, \qquad
\left\{    y_i^{\alpha},z_j^{\beta}   \right\}= \delta_{ij} \, z_j^{\gamma}, \qquad
\left\{    z_i^{\al},z_j^{\be}   \right\}= 0 , \qquad \alpha,\beta,\gamma= 1,2,3. \label{e3}
\eeq

\section{The rational Lagrange chain}

As we have shown in the previous section
the $\mathfrak{e}(3)$ Lagrange chain is derived from the following $2 \times 2$ Lax matrix
\beq
\CL(\l)= w \, \sigma^3 +  \sum_{i=1}^N   \sum_{\alpha=1}^3
\sig^{\alpha} \left[\frac{y_i^{\al}}{\l - \l_i}
+  \frac{z_i^{\al}}{(\l - \l_i)^2 } \right].\label{lax}
\eeq
The parameter 
$w$ denotes the intensity of an external field, 
taken as uniform (along the chain) and constant (in time). 

Recall that a generic element $\xi \in \mathfrak{su}(2)$ may be written as
$$
\xi = \half \left(\begin{array}{cc}
{\rm{i}} \, \xi^{3} &  {\rm{i}} \,\xi^{1}+ \xi^2 \\
{\rm{i}} \,\xi^{1}- \xi^2 & -\, {\rm{i}} \, \xi^{3}
\end{array}\right).
$$
If we associate to this matrix the vector
$$
\xi =(\xi^1,\xi^2,\xi^3)^T \in \mathbb{R}^3
$$
then it is easy to verify that this correspondence is an isomorphism between
$\mathfrak{su}(2)$ and the Lie algebra $(\mathbb{R}^3, [\cdot,\cdot])$, where
the Lie bracket
$[\cdot,\cdot]$ is realized with the vector product.

Let us now fix the following notation: 
$$
{\bf{Y}}_i \doteq (y_i^1,y_i^2,y_i^3)^T \in \mathbb{R}^3, \qquad 
{\bf{Z}}_i \doteq (z_i^1,z_i^2,z_i^3)^T \in \mathbb{R}^3, \qquad i=1,...N,$$
$$
{\bf{y}}^\alpha \doteq (y^\alpha_1,...,y^\alpha_N)^T \in \mathbb{R}^N, \qquad
{\bf{z}}^\alpha \doteq (z^\alpha_1,...,z^\alpha_N)^T \in \mathbb{R}^N, \qquad \alpha=1,2,3.$$
Namely ${\bf{Y}}_i$ denotes the $i$-th angular momentum and 
${\bf{Z}}_i$ denotes the vector pointing 
from the fixed point to the center of mass of the $i$-th top;
${\bf{y}}^\alpha$ and ${\bf{z }}^\alpha$ denote respectively
the sets of $N$ angular momenta and position vectors with $\alpha$ fixed.

The Lie--Poisson brackets (\ref{e3}) have $2N$ Casimir functions:
\beq
C_i^{(1)} = \langle{\bf{Y}}_i, {\bf{Z}}_i \rangle \doteq 
\sum_{\alpha=1}^3 y^\al_i \, z^\al_i, \qquad
C_i^{(2)} = \langle{\bf{Z}}_i,{\bf{Z}}_i \rangle \doteq
\sum_{\alpha=1}^3 (z^\al_i)^2 \qquad i=1,...,N. \nonumber
\eeq 
Fixing their values one gets a $2N$-dimensional symplectic leaf
$$
\mathcal{O} 
\doteq \left\{    ({\bf{Y}}_i,{\bf{Z}}_i) \in
\mathbb{R}^3 \times \mathbb{R}^3, i=1,..,N \;|
\; C_i^{(1)} = \ell_i,
\; C_i^{(2)}= 1 \right\}. \label{leaf}
$$

As we have shown in \cite{MPR} the
Lax matrix (\ref{lax}) satisfies the linear $r$-matrix algebra
\beq
\left\{\CL(\l) \otimes \mathds{1}, \mathds{1} \otimes \CL(\mu) \right\}-
\left[r(\l - \mu), \CL(\l) \otimes \mathds{1} + \mathds{1} \otimes \CL(\mu) \right]=0,
\label{333}
\eeq
where $\mathds{1}$ is the $2 \times 2$ identity matrix and the $r$-matrix is given by
$$
r(\la)=\frac{{\rm i}}{ \l} \sum_{\al=1}^3  \sig^{\al} \otimes \sig^{\al}.
\label{666}
$$

The spectral curve $\Gamma$,
\beq
\Gamma: \quad \det( \CL(\l) -\mu \mathds{1})=0, \label{gamma}
\eeq
is an hyperelliptic curve of genus $g=2N-1$, reading
$$
-\mu^2 =
w^2+
\sum_{i=1}^N \frac{R_i}{\l-\l_i}+
\frac{S_i}{(\l-\l_i)^2}+
\frac{C_i^{(1)}}{(\l-\l_i)^3}+
\frac{C_i^{(2)}}{(\l-\l_i)^4},\nonumber
$$
with the Hamiltonians $R_i$ and $S_i$ given by:
\begin{eqnarray}
R_i &=&\langle{\bf{W}}, {\bf{Y}}_i \rangle + \sum_{k \neq i }^N
\left(
\frac{ \langle{\bf{Y}}_i, {\bf{Y}}_k \rangle}{\l_i - \l_k}+ 
\frac{ \langle{\bf{Y}}_i, {\bf{Z}}_k \rangle  -  \langle{\bf{Y}}_k, {\bf{Z}}_i \rangle }
{(\l_i - \l_k)^2}
 - 2 \frac{ \langle{\bf{Z}}_i, {\bf{Z}}_k \rangle }{(\l_i - \l_k)^3}
\right), \nonumber \\
&& \nonumber \\
S_i &=&\langle{\bf{W}}, {\bf{Z}}_i\rangle + \frac{\langle{\bf{Y}}_i, {\bf{Y}}_i \rangle}{2} +
\sum_{k \neq i }^N
\left(
\frac{ \langle{\bf{Y}}_k, {\bf{Z}}_i \rangle }{\l_i - \l_k}+
 \frac{ \langle{\bf{Z}}_i, {\bf{Z}}_k \rangle }{(\l_i - \l_k)^2}
\right), \nonumber 
\end{eqnarray}
where ${\bf{W}} \doteq (0,0,w)^T \in \mathbb{R}^3$ is the external field vector.
These are integrals of motion of the rational Lagrange chain, 
which are Poisson commuting due to (\ref{333}):
\beq
\left\{R_i,R_j\right\}=\left\{S_i,S_j\right\}=\left\{R_i,S_j\right\}=0, \quad i,j=1,...,N. \nonumber
\eeq
Notice that the linear integral
\beq
\sum_{i=1}^N R_i= \sum_{i=1}^N \langle{\bf{W}},{\bf{Y}}_i \rangle, \nonumber
\eeq
yields the third component of the total angular momentum of the chain.

We can bring the curve $\Gamma$ into the canonical form by the scaling
\beq
\mu \longmapsto \hat \mu = \mu \prod_{i=1}^{N} (\l -\l_i)^2.\nonumber
\eeq
The equation of the spectral curve becomes
\bea
-\hat \mu^2 & =& \left[
w^2+
\sum_{i=1}^N \frac{R_i}{\l-\l_i}+
\frac{S_i}{(\l-\l_i)^2}+
\frac{C_i^{(1)}}{(\l-\l_i)^3}+
\frac{C_i^{(2)}}{(\l-\l_i)^4}  \right] \, \prod_{i=1}^{N} (\l -\l_i)^4 = \nonumber \\
&&  \nonumber \\
&=& w^2 \l^{4N}+s_1 \l^{4N-1} +s_2 \l^{4N-2} + ...+ s_{4N}, \nonumber 
\eea
where the coefficients $s_j$, $j=1,...,4N$ are linear combinations of the Hamiltonians
and the Casimir functions.

In the following we will use complex conjugated generators 
$$
y^\pm_i \doteq y^1_i \, \pm \,  {\rm{i}} \, y^2_i \qquad i=1,...,N,
$$
which have the brackets:
$$
\left\{    y_i^{3},y_j^{\pm}   \right\}= \mp \delta_{ij} \; y_j^{\pm}, \qquad
\left\{    y_i^{+},y_j^{-}   \right\}= -2 {\rm{i}} \; \delta_{ij} \;y_j^{3},
$$
$$
\left\{    y_i^{3},z_j^{\pm}   \right\}= \left\{    z_i^{3},y_j^{\pm}   \right\}=
\mp {\rm{i}} \; \delta_{ij} \; z_j^{\pm}, \qquad 
\left\{    y_i^{+},z_j^{-}   \right\}= \left\{    z_i^{+},y_j^{-}   \right\}= 
-2  {\rm{i}} \; \delta_{ij} \; z_j^{3},
$$
$$
\left\{ y_i^{\be},z_j^{\be}   \right\}= 0, \qquad
\left\{ z_i^{\al},z_j^{\be}   \right\}= 0 , \qquad \alpha,\beta,\gamma= \pm,3. 
$$
In terms of these generators the Lax matrix (\ref{lax}) has the following explicit form
\begin{equation}
\CL(\l)= \CW 
+ \sum_{i=1}^{N} \left[ \frac{\CY_i}{\l-\l_{i}}+
\frac{\CZ_i }{(\l-\l_{i})^2} \right], \label{laxx}
\end{equation}
where $\CW \doteq w \, \sigma^3 \in \mathfrak{su}(2)$ and 
$$
\CY_i \doteq \frac{{\rm{i}}}{2}\left(\begin{array}{cc}
 y_{i}^{3} &  y_{i}^{-} \\
 y_{i}^{+} & - y_{i}^{3}
\end{array}\right) \, \in \mathfrak{su}(2), \qquad
\CZ_i \doteq \frac{{\rm{i}}}{2} \left(\begin{array}{cc}
 z_{i}^{3} & z_{i}^{-} \\
z_{i}^{+} & -z_{i}^{3}
\end{array}\right) \, \in \mathfrak{su}(2) .
$$

\section{A Lax formulation}

A possible choice for a physical Hamiltonian describing the dynamics of the model is the following
one:
\beq
\CH= \sum_{i=1}^N ( \l_i \, R_i + S_i)=
 \sum_{i=1}^N \langle{\bf{W}}, \l_i {\bf{Y}}_i + {\bf{Z}}_i\rangle+
\half \sum_{i,k=1}^N \langle{\bf{Y}}_i, {\bf{Y}}_k \rangle. \label{H!}
\eeq
Let us remark that if $N=1$ the Lagrange chain degenerates into the well-known 
$\mathfrak{e}(3)$ symmetric Lagrange top, whose Hamiltonians are given by
$y^3$ and $[(y^1)^2+(y^2)^2+(y^3)^2]/2 + w  z^3$ \cite{Au,RSTS, KPR}.

In the case of the Hamiltonian (\ref{H!}) the equations of motion are given by
\beq
\left\{ \begin{array}{ll}
\dot {\bf{Y}}_i = {\bf{W}} \wedge  {\bf{Z}}_i +
\left( \l_i {\bf{W}} + \sum_{k=1}^N {\bf{Y}}_k \right)
 \wedge {\bf{Y}}_i \, , \\
\dot {\bf{Z}}_i = \left( \l_i {\bf{W}} + \sum_{k =1}^N {\bf{Y}}_k \right)
 \wedge {\bf{Z}}_i \, ,
\end{array}\right. \qquad i=1,...,N, \label{eq}
\eeq
where $\wedge$ is the standard vector product
in $\mathbb{R}^3$. Note again that if $N=1$ equations (\ref{eq}) coincide with the
equations of motion of the symmetric Lagrange top \cite{Au,KPR,BS} provided that
$\l_1=0$.

We can write the equation of motion in the following form:
\beq
\left\{ \begin{array}{ll}
\dot {\CY}_i = \left[{\CW} ,{\CZ}_i\right] +
\left[ \l_i {\CW} + \sum_{k =1}^N {\CY}_k ,
 {\CY}_i\right] \, , \\
\dot {\CZ}_i = \left[ \l_i {\CW} + \sum_{k =1}^N {\CY}_k ,
 {\CZ}_i \,\right] ,
\end{array}\right. \qquad i=1,...,N. \label{eq2}
\eeq

One has the following statement:

\begin{prop}

The Lax representation for equations (\ref{eq2}) is given by
$$
\dot \CL (\l) = \left[  \CL(\l), \CM (\l)  \right]
$$
where the two matrices from the loop algebra $\mathfrak{su}(2)[\l]$ are the following ones:
\bea
&& \CL (\l) = \CW
+ \sum_{i=1}^{N} \left[ \frac{\CY_i}{\l-\l_{i}}+
\frac{\CZ_i }{(\l-\l_{i})^2} \right], \nonumber \\
&& \CM (\l) = \sum_{i=1}^{N} \frac{1}{\l-\l_{i}} \left[
\l_i \, \CY_i + \frac{\l \, \CZ_i }{\l-\l_{i}}
\right]. \nonumber 
\eea
\end{prop}

{\bf{Proof:}} A direct calculation.

\endpf

Let us notice that in the case $N=1$ we recover 
the well--known Lax representation for the symmetric Lagrange top \cite{Au,RSTS}
provided that $\l_1=0$:
$$
\dot \CL (\l) = \left[  \CL(\l), \CM (\l)  \right]
$$
where 
$$\CL (\l) = \CW
+  \frac{\CY}{\l}+
\frac{\CZ }{\l^2} , \qquad \CM (\l) = \frac{\CZ}{\l}. \nonumber 
$$

\section{Separation of variables}

In this section we construct the simplest separation of variables
for the Lagrange chain with the Lax matrix (\ref{laxx}).
The details of the approach can be found in \cite{Skl38, KNS}.

The basic separation has only $N-1$ pairs of separation variables
belonging to  the spectral curve $\Gamma$ (\ref{gamma}). 
It corresponds to the standard normalization vector
$\alpha_0=(1,0)$ and it is defined by the equations
\begin{equation}
(1,0) \; (\CL(u)- v \mathds{1})^\wedge=0, \qquad u,v \in \mathbb{C}\label{ff}
\end{equation}
where $(\cdot)^\wedge$ denotes the adjoint matrix. Equation (\ref{ff}) is the equation for
the poles $(u_k,v_k)$
of the Baker-Akhiezer function $\Psi$, which is defined as a properly normalized eigenfunction
of the Lax matrix:
\begin{equation}
\CL(u) \, \Psi = v \, \Psi, \qquad (u,v) \in \Gamma, \nonumber
\end{equation}
\begin{equation}
\langle \alpha_0 ,\Psi \rangle  =1. \nonumber
\end{equation}
It is easy to see that the
equation (\ref{ff}) gives the following separation variables:
\begin{equation}
\CL_{12}(u_k)=0, \qquad v_k=-\CL_{11}(u_k), \nonumber
\end{equation}
where $\CL_{jk}(\l)$ denotes the entry $jk$ of the Lax matrix $\CL(\l)$.
Explicitly, the first $N-1$ components $u_k$ of the separation variables are defined as 
zeros of the element $\CL_{12}$ of the Lax matrix (\ref{laxx}):
\beq
\CL_{12}(u_k)= {\rm{i}} \sum_{i=1}^N \left[
\frac{y^-_{i}}{u_k - \l_i}+\frac{z^-_{i}}{(u_k - \l_i)^2} \right] =0 \qquad k=1,...,N-1,\label{11}
\eeq
while the second components are the values of $-\CL_{11}(u)$ in those zeros:
\beq
v_k= -\CL_{11}(u_k) =- {\rm{i}} \, w- {\rm{i}} \sum_{i=1}^N \left[
\frac{y^3_{i}}{u_k - \l_i}+\frac{z^3_{i}}{(u_k - \l_i)^2} \right] \qquad k=1,...,N-1.
\label{22}
\eeq
Let us show the canonicity of these separation variables. We obtain
$$
0 \equiv \left\{\CL_{12}(u_k),v_l\right\}= \frac{d \CL_{12}(u_k)}{d u_k}
\left\{u_k,v_l\right\} + \left\{\CL_{12}(u),v_l\right\}_{u=u_k}.
$$
Using the linear $r$-matrix Poisson algebra (\ref{333}) we obtain
$$
\left\{u_k,v_l\right\}= - \left( \frac{d \CL_{12}(u_k)}{d u_k}  \right)^{-1}
\left\{\CL_{12}(u),v_l\right\}_{u=u_k}= \delta_{kl} \qquad k,l=1,...,N-1.
$$
For the completeness, one has to add an extra pair of canonical variables which Poisson commute with all
separation variables, in order to make the total number of new Darboux variables equal twice the number
of degrees of freedom. This pair is taken from the asymptotics of the elements $\CL_{11}(u)$ and
$\CL_{12}(u)$:
$$
\CL_{11}(u)= {\rm{i}}\, w + {\rm{i}} \sum_{i=1}^N \frac{y^3_{i}}{u} + O \left(\frac{1}{u^2}  \right), \quad
\CL_{12}(u)= {\rm{i}} 
\sum_{i=1}^N \frac{y^-_{i}}{u} + O \left(\frac{1}{u^2}  \right) \qquad u \rightarrow \infty.
$$
Thus we can choose as the last pair of separation variables the following ones:
$$
u_N= {\rm{i}} \sum_{i=1}^N y^-_{i}, \qquad v_N= - \frac{\sum_{i=1}^N y^3_{i}}{\sum_{i=1}^N y^-_{i}}.
$$
Indeed, it is easy to check that $u_N,v_N$ commute with the separation variables defined in 
(\ref{11}) and (\ref{22}).

\section{B\"acklund transformations}

In this paper, following the approach of \cite{KS5,KV00}, we look at the 
B\"acklund transformations (BTs) for finite-dimensional
(Liouville)
integrable systems as special canonical transformations. 
Such BTs are defined as symplectic, or
more
generally Poisson, integrable maps which are explicit maps (rather than implicit
multivalued correspondences) and which can be viewed as time discretizations of
particular
continuous flows. 

The most characteristic properties of such maps are: 
\begin{enumerate}
\item a BT preserves the
same set of integrals of motion as does the continuous flow which it discretizes;
\item it 
depends on a B\"acklund parameter $\eta$ that specifies the corresponding
shift on a Jacobian or on a generalized Jacobian \cite{KV00};
\item a spectrality
property holds with respect
to $\eta$ and to the conjugate variable $\mu$, which means that the point
$(\eta,\mu)$ belongs
to the spectral curve \cite{KS5,KV00}.
\end{enumerate}
Explicitness makes these maps purely iterative, while the importance of the parameter
$\eta$
is that it allows for an adjustable discrete time step. The spectrality property is
related with the simplecticity of the map \cite{KV00}.

\subsection{One--point BTs}

A one-point B\"acklund transformation 
for the $\mathfrak{e}(3)$ rational Lagrange chain can be defined
as the following similarity transform on the Lax matrix $\CL(\l)$ (\ref{laxx}):
$$
\CL(\l)  \longmapsto \CM(\l;\eta) \,  \CL(\l) \, \CM^{-1}(\l,\eta) \qquad \forall ~\l \in \mathbb{C},
$$
with some generally non-degenerate $2 \times2$ matrix $\CM(\l,\eta)$, simply because a BT should
preserve
the spectrum of $\CL(\l)$. The parameter $\eta \in \mathbb{C}$ is called a B\"acklund parameter of
the transformation. We use $~ \til {}$ -notations for the updated variables, so that
$$
\til \CL(\l)= \CW
+ \sum_{i=1}^{N} \left[ \frac{\til \CY_i}{\l-\l_{i}}+
\frac{\til \CZ_i }{(\l-\l_{i})^2} \right],
$$
where
$$
\til \CY_i \doteq \frac{{\rm{i}}}{2} \left(\begin{array}{cc}
\til y_{i}^{3} & \til y_{i}^{-} \\
\til y_{i}^{+} & -\til y_{i}^{3}
\end{array}\right), \qquad
\til \CZ_i \doteq \frac{{\rm{i}}}{2} \left(\begin{array}{cc}
\til z_{i}^{3} & \til z_{i}^{-} \\
\til z_{i}^{+} & -\til z_{i}^{3}
\end{array}\right).
$$
We are looking for a Poisson map that intertwines two Lax matrices
$\CL(\l)$ and $\til
\CL(\l)$:
\begin{equation}
\CM(\l;\eta) \, \CL(\l)=\til \CL(\l) \, \CM(\l;\eta) \qquad \forall ~\l \in \mathbb{C}, \quad
\eta \in \mathbb{C}.
\label{rr}
\end{equation}
Let us take
\begin{equation}
\CM(\l;\eta)=\left(\begin{array}{cc}
\l-\eta +p \, q& p \\
q & 1
\end{array}\right),\qquad \det \CM(\l;\eta) =\l - \eta.
\label{ansatz}
\end{equation}
We stress that the number of zeros of $\det \CM$ is the number of essential B\"acklund
parameters. Here the variables $p$ and $q$ are indeterminate dynamical variables.
The ansatz (\ref{ansatz}) for the matrix $\CM$ comes from the simplest
$L$-operator of the quadratic $r$-matrix algebra
with the same $r$-matrix of the model \cite{HKR,KPR}.

Comparing the asymptotics in $u\rightarrow\infty$ in both sides of (\ref{rr}) we readily get
\beq
\qquad p=\frac{1}{2w}\sum_{i=1}^{N} y^-_j, \qquad
q=\frac{1}{2w} \sum_{i=1}^{N} \til y^+_j.  \label{hh}
\eeq
If we want an explicit single-valued map from $\CL(\l)$ to $\til \CL(\l)$
we must express $\CM(\l;\eta)$, and therefore $p$ and $q$ in term of the old variables. To
solve this problem
we use the spectrality of the BT \cite{KS5,KV00}. Equation (\ref{rr}) defines a map $\mathcal{B}_{P}$
parametrized by the point
point $P=(\eta,\mu) \in \Gamma$. Notice that there are two points on
$\Gamma$,
$P=(\eta,\mu)$ and $Q=(\eta,-\mu)$, corresponding to the same $\l$ and sitting
one above
the other because of the hyperelliptic involution:
\begin{equation}
(\eta,\mu) \in \Gamma: \qquad \det(\CL(\eta)
-\mu \mathds{1})=0. \nonumber
\end{equation}
This spectrality property, used as a new datum, produces
the formula
\beq
q=\frac{\CL_{11}(\eta)-\mu}{\CL_{12}(\eta)}=-\frac{\CL_{21}(\eta)}{\CL_{11}(\eta)+\mu}\,.
\label{hhh}
\eeq

Now the equation (\ref{rr}) gives an integrable Poisson map from $\CL(\l)$ to $\tilde \CL(\l)$. 
We recall here the notation
${\bf{y}}^\alpha \doteq (y^\alpha_1,...,y^\alpha_N)^T \in \mathbb{R}^N$ and
${\bf{z}}^\alpha \doteq (z^\alpha_1,...,z^\alpha_N)^T \in \mathbb{R}^N$, with $\alpha=\pm,3$.

The following statement shows how the one-point BT can be written in a symplectic form
through a generating function.

\begin{prop} The one--point BT for the rational Lagrange chain is defined by
\begin{eqnarray}
z^3_i &=& {\rm{i}} \,  z^-_i \,\frac{\partial
F_\eta ({\bf{z}}^-, {\bf{y}}^- \mid \til {\bf{z}}^+, \til {\bf{y}}^+)}
{\partial y^-_i}\,, \nonumber \\
y^3_i &=& {\rm{i}}\,z^-_i \,\frac{\partial
F_\eta ({\bf{z}}^-, {\bf{y}}^- \mid \til {\bf{z}}^+, \til {\bf{y}}^+)}
{\partial z^-_i}+
{\rm{i}}\,y^-_i \,\frac{\partial
F_\eta ({\bf{z}}^-, {\bf{y}}^- \mid \til {\bf{z}}^+, \til {\bf{y}}^+)}
{\partial y^-_i}\,,\nonumber\\
\til z^3_i &=&
{\rm{i}}\, \tilde z^+_i \,\frac{\partial
F_\eta ({\bf{z}}^-, {\bf{y}}^- \mid \til {\bf{z}}^+, \til {\bf{y}}^+)}
{\partial \til y^+_i}\,,\nonumber\\
\til y^3_i &=&
{\rm{i}}\, \tilde z^+_i \,\frac{\partial
F_\eta ({\bf{z}}^-, {\bf{y}}^- \mid \til {\bf{z}}^+, \til {\bf{y}}^+)}
{\partial \til z^+_i}+
{\rm{i}}\, \til y^+_i\, \frac{\partial
F_\eta ({\bf{z}}^-, {\bf{y}}^- \mid \til {\bf{z}}^+, \til {\bf{y}}^+)}
{\partial \til y^+_i}\,,
\label{gf}
\end{eqnarray}
with
\bea
F_\eta ({\bf{z}}^-, {\bf{y}}^- \mid \til {\bf{z}}^+, \til {\bf{y}}^+)&=&
-\frac{{\rm{i}}}{2w} \sum_{i,j=1}^{N} y^-_i \til y^+_j - {\rm{i}} \sum_{i=1}^{N}
 k_i \left( \frac{y^-_i}{z^-_i}+
\frac{\til y^+_i}{\til z^+_i} -\frac{1}{\eta -\l_i} \right) + \nonumber \\ 
&&+ \, {\rm{i}} \log \prod_{i=1}^{N}
\left(\frac{1+k_i}{1-k_i}\right)^{\ell_i} - {\rm{i}}\, \, w \, N \, \eta \, , \label{zzz}
\eea
where
$$
k_i^2 =1+ (\eta - \l_i) z_i^-\, \til z^+_i.
$$
\end{prop}

{\bf{Proof:}} 
First, because the Casimir functions $ C_i^{(1)},C_i^{(2)}, i=1,...,N$ do not change under the map,
namely
$$
y^3_i z^3_i + \frac{1}{2} \left( y^-_i z^+_i + y^+_i z^-_i \right)=
\til y^3_i \til z^3_i + \frac{1}{2} \left( \til y^-_i \til z^+_i + \til y^+_i \til z^-_i \right)
=\ell_i,
$$
$$
(z^3_i)^2 + z^-_i z^+_i=(\til z^3_i)^2 + \til z^-_i \til z^+_i=1,
$$
we can exclude $4N$ variables $z^+_i,y^+_i$ and  $\til z_i^-,\til y^-_i$, with $i=1,...,N$,
using the following substitutions:
$$
z^+_i=\frac{1-(z^{3}_i)^{2}}{z^-_i}\,, \qquad \qquad y^+_i=\frac{2 \ell_i}{z^-_i}-
\frac{2 y^3_i z^3_i}{z^-_i}-
\frac{y^-_i}{(z^-_i)^{2}}\,[1-(z^3_i)^{2}]\,,
$$
$$
\til z^-_i=\frac{1-(\til z^{3}_i)^{2}}{\til z^+_i}\,, \qquad \qquad
\til y^+_i=\frac{2 \ell_i}{\til z^+_i}-\frac{2  \til y^3_i \til z^3_i}
{\til z^+_i}-
\frac{\til  y^+_i}{(\til z^+_i)^{2}}\,[1-(\til z^3_i)^{2}].
$$
Now we have only $4N +4N$ (old and new) independent variables:
$z^-_i,z^3_i$, $y^-_i,y^3_i$ and  $\til z^+_i,\til z^3_i$, $\til y^+_i,\til y^3_i, i=1,...,N$.

The map (\ref{rr}) explicitly reads
\bea
&& \til  \CL_{11}(\l) = 
\frac{(\l -\eta +2\,p \, q)[\CL_{11}(\l) -q \, \CL_{12}(\l)]+p \, \CL_{21}(\l)}{\l -\eta}, \nonumber \\
&& \til  \CL_{12}(\l) = 
\frac{(\l -\eta +2\,p\,q)^2 \CL_{12}(\l) -2\,p\,(\l -\eta +2 \,p\,q) \CL_{11}(\l) -p^2 \CL_{21}(\l)}
{\l -\eta}, \nonumber \\
&& \tilde  \CL_{21}(\l) = 
\frac{\CL_{21}(\l) +2\,q \, \CL_{11}(\l) -q^2 \CL_{12}(\l)}
{\l -\eta} \nonumber
\eea
Equating residues at $\l=\l_i$ in both sides of the above equations we obtain, after a straightforward computation,
\bea
z^3_i &=& \frac{z^-_i}{2w} \sum_{j=1}^{N} \til y^+_j
 +k_i, \nonumber \\
y^3_i &=&\frac{\ell_i}{k_i}+
\frac{\eta-\lambda_i}{2k_i}\,\left(\til z^+_i y^-_i   +
z^-_i \til y^+_i \right) -
\frac{z^-_i\til z^+_i}{2k_i}+\frac{y^-_i}{2w}\sum_{j=1}^{N} \til y^+_j\,,\nonumber\\
\til z^3_i &=& \frac{\til x_+}{2w} \sum_{j=1}^{N} y^-_j +k_i,\nonumber\\
\til y^3_i&=& \frac{\ell_i}{k_i}+
\frac{\eta-\lambda_i}{2k_i}\,\left(\til z^+_i  y^-_i  +
z^-_i \til y^+_i \right) -
\frac{z^-_i\til z^+_i}{2k_i}+\frac{\til y^+_i}{2w}\sum_{j=1}^{N} y^-_j\,. \label{zz}\\
\nonumber
\eea
where
$$
k_i^2 =1+ (\eta - \l_i) z_i^- \til z^+_i.
$$
It is now easy to check that the function 
$F_\eta ({\bf{z}}^-, {\bf{y}}^- \mid \til {\bf{z}}^+, \til {\bf{y}}^+)$ (\ref{zzz})
satisfies equations (\ref{gf}).
\endpf

The spectrality property of a B\"acklund transformation means that the two coordinates $\eta$
and $\mu$ of the point $P \in \Gamma$ parametrizing the map are conjugated variables, namely
$$
\mu = -\frac{\partial F}{ \partial \eta}\,,
$$
where $F$ is the generating function of the BT.

We now show the spectrality property for the one-point constructed BT. Using the 
equations (\ref{hh}), (\ref{hhh}), (\ref{zzz}) and (\ref{zz}) we obtain
\bea
\mu&=&\CL_{11}(\eta)-\left(
\frac{1}{2w} \sum_{i=1}^{N} \til y^+_j \right)
 \, \CL_{12}(\eta)= \nonumber \\
 &=& {\rm{i}} \, \sum_{i=1}^N
\frac{1}{k_i} \left[
\frac{1}{(\eta - \l_i)^2} + \frac{\ell_i}{\eta - \l_i}+
\frac{z^-_i \til z^+_i}{2} \left(
\frac{y^-_i}{z^-_i} + \frac{\til y^+_i}{\til z^+_i} + \frac{1}{\eta - \l_i}
\right) \right] + {\rm{i}} \, w \, N= \nonumber  \\
&=& -\frac{\partial F_\eta ({\bf{z}}^-, {\bf{y}}^- \mid \til {\bf{z}}^+, \til {\bf{y}}^+)}
{\partial \eta}\,.
\nonumber
\end{eqnarray}

Notice that the above one-point BT is a complex map, so it is a non-physical
B\"acklund transformation. In order to obtain a physical map we will construct a two--point
BT in the next section.

\subsection{Two--point BTs}

According to \cite{HKR,KPR}, we now construct a composite map which is a product of the map
$\mathcal{B}_{P_1} \equiv \mathcal{B}_{(\eta_1,\mu_1)}$ and
$\mathcal{B}_{Q_2} \equiv \mathcal{B}_{(\eta_2,-\mu_2)}$:
\begin{equation}
\mathcal{B}_{P_1,Q_2}=\mathcal{B}_{Q_2} \circ \mathcal{B}_{P_1}:~
\CL(\l) \stackrel{\mathcal{B}_{P_1}} {\longmapsto} \, \til \CL(\l)
\stackrel{\mathcal{B}_{Q_2}} {\longmapsto} \; \stackrel{\approx}{\CL}(\l). \nonumber
\end{equation}
The two maps are inverse to each other when $\eta_1=\eta_2$ and $\mu_1=\mu_2$.
This  two-point BT 
for the rational Lagrange chain is defined by the following ``discrete-time'' Lax equation:
\beq
\CM(\l;\eta_1,\eta_2) \, \CL(\l)=\stackrel{\approx}{\CL}(\l) \,\CM(\l;\eta_1,\eta_2) 
\qquad \forall \l\in \mathbb{C}, \qquad 
\eta_1,\eta_2 \in \mathbb{C},
\label{128}\end{equation}
where the matrix $\CM(\l;\eta_1,\eta_2)$ is \cite{HKR,KPR}
\beq
\CM(\l;\eta_1,\eta_2)=\begin{pmatrix}\l-\eta_1+x \, X
&X\cr -x^2\,X+(\eta_1-\eta_2)\, x&\l-\eta_2-x\,X\end{pmatrix},
\label{mm} 
\eeq
$$
\det \CM(\l;\eta_1,\eta_2) =(\l - \eta_1)(\l - \eta_2).
$$
The spectrality property with respect to two fixed points $(\eta_1,\mu_1)\in \Gamma$
and $(\eta_2,\mu_2)\in \Gamma$ give
\bea
x&=&\frac{\CL_{11}(\eta_1)-\mu_1}{\CL_{12}(\eta_1)}=- \frac{\CL_{21}(\eta_1)}{\CL_{11}(\eta_1)+ \mu_1}
=\frac{\stackrel{\approx}{\CL}_{11}(\eta_2)-\mu_2}
{\stackrel{\approx}{\CL}_{12}(\eta_2)}=-\frac{\stackrel{\approx}{\CL}_{21}(\eta_2)}
{\stackrel{\approx}{\CL}_{11}(\eta_2)
+\mu_2}= \nonumber \\
&=&\frac{1}{2w} \sum_{i=1}^{N} \til y^+_j. \label{x} \\
X&=&\frac{(\eta_2-\eta_1)\CL_{12}(\eta_1)\CL_{12}(\eta_2)}
{\CL_{12}(\eta_1)(\CL_{11}(\eta_2)+\mu_2)-\CL_{12}(\eta_2)(\CL_{11}(\eta_1)-\mu_1)}=
\nonumber \\
&=&\frac{(\eta_1-\eta_2)(\CL_{11}(\eta_1)+\mu_1)(\CL_{11}(\eta_2)-\mu_2)}
{(\CL_{11}(\eta_1)+\mu_1)\CL_{21}(\eta_2)-(\CL_{11}(\eta_2)-\mu_2)\CL_{21}(\eta_1)}=\nonumber\\
&=&\frac{(\eta_2-\eta_1)\stackrel{\approx}{\CL}_{12}(\eta_1)
\stackrel{\approx}{\CL}_{12}(\eta_2)}
{\stackrel{\approx}{\CL}_{12}(\eta_2)(\stackrel{\approx}{\CL}_{11}(\eta_1)+\mu_1)-
\stackrel{\approx}{\CL}_{12}(\eta_1)(\stackrel{\approx}{\CL}_{11}(\eta_2)-\mu_2)}= \nonumber\\
&=&\frac{(\eta_1-\eta_2)(\stackrel{\approx}{\CL}_{11}(\eta_1)-\mu_1)
(\stackrel{\approx}{\CL}_{11}(\eta_2)+\mu_2)}
{(\stackrel{\approx}{\CL}_{11}(\eta_2)+\mu_2)\stackrel{\approx}{\CL}_{21}(\eta_1)-
(\stackrel{\approx}{\CL}_{11}(\eta_1)-\mu_1)\stackrel{\approx}{\CL}_{21}(\eta_2)}= \nonumber\\
&=& \frac{1}{2w} \sum_{i=1}^{N} \left( y^-_j   - \stackrel{\approx}{y}^+_j \right) \label{X}.
\end{eqnarray}
Now we have two 
B\"acklund parameters $\eta_1,\eta_2 \in \mathbb{C}$. The above formulae
give several equivalent
expressions for the variables $x$ and $X$ since the points
$(\eta_1,\mu_1)$ and $(\eta_2,\mu_2)$ belong to the spectral curve $\Gamma$,
i.e., are bound
by the following relations
$$
\mu_k^2=\CL_{11}^2(\eta_k)+\CL_{12}(\eta_k)\CL_{21}(\eta_k)=
\stackrel{\approx}{\CL}_{11}^2(\eta_k)+\stackrel{\approx}{\CL}_{12}(\eta_k)
\stackrel{\approx}{\CL}_{21}(\eta_k),
\qquad k=1,2.
$$

Together with (\ref{x}) and (\ref{X}), the 
 formula (\ref{128}) give an explicit two-point Poisson integrable map from $\CL(\l)$ to
$\stackrel{\approx}{\CL}(\l)$ (as well as its inverse, i.e the map from 
$\stackrel{\approx}{\CL}(\l)$ to $\CL(\l)$).
The map is parametrized by two points $\mathcal{B}_{P_1} \equiv
\mathcal{B}_{(\eta_1,\mu_1)}$ and
$\mathcal{B}_{Q_2} \equiv \mathcal{B}_{(\eta_2,-\mu_2)}$.

Obviously, when $\eta_1=\eta_2$ (and $\mu_1=\mu_2$) the map turns
into an identity map. 
We want to stress the fact that the two--point BT sends real variables
to real variables provided \cite{KPR}
$$
\eta_1=\bar\eta_2 \doteq \eta \in \mathbb{C}. \label{gg}
$$
Therefore,  the two-point map leads to a physical B\"acklund transformation
with two real parameters.

\section{Concluding remarks}

In the present paper we investigated a new classical integrable system, which we called ``rational
Lagrange chain''. It consists of a long--range homogeneous chain where the local variables are
the generators of the direct sum of $N$ $\mathfrak{e}(3)$ interacting Lagrange tops. In our
previous paper \cite{MPR} we have shown that this model can be derived from the $N$ sites
$\mathfrak{su}(2)$ rational Gaudin models. Moreover
this construction can be generalized to
trigonometric and elliptic solutions of the classical Yang--Baxter
equation and to a generic finite--dimensional simple Lie algebra.

We obtained a Lax representation for the system which naturally reduces 
to the well--known Lax pair of the Lagrange top in the one--body case.

Following the approach proposed by V.B. Kuznetsov and E.K. Sklyanin for the construction
of B\"acklund transformations for finite--dimensional systems \cite{KS5,KV00}, we 
obtained one- and two--point integrable symplectic maps for the Lagrange chain. Our explicit maps
are the natural generalization to the $N$ bodies system of the BTs for the symmetric
Lagrange top \cite{KPR}.

\section*{Acknowledgments}

Two of us (MP and OR) would like to thank Vadim B. Kuznetsov for having suggested the
way to construct BTs for finite dimensional systems.
It is also a pleasure to acknowledge interesting discussions with Yuri B. Suris.



\begin{thebibliography}{40}

\bibitem {Au} M. Audin, {\it{Spinning tops}}, Cambridge University Press (1996).

\bibitem {BS} A.I. Bobenko and Yu. B. Suris, {\it{Discrete time Lagrangian mechanics on
Lie groups, with an application to the Lagrange top}}, 
Commun. Math. Phys. {\bf 204} (1999) 147--188.

\bibitem {G1} M. Gaudin, {\it{Diagonalisation d'une classe d' hamiltoniens de spin}},
J. de Physique {\bf 37} (1976) 1087--1098.

\bibitem {G2} M. Gaudin, {\it{La fonction d' onde de Bethe}}, Masson, Parigi (1983).

\bibitem{HKR} A.N.W.~Hone, V.B.~Kuznetsov and O.~Ragnisco,
{\it B\"{a}cklund transformations for the $sl(2)$ Gaudin magnet},
Journal of  Physics A {\bf 34} (2001), 2477--2490.

\bibitem{MPR} F. Musso, M. Petrera, O.~Ragnisco,
{\it Algebraic extensions of Gaudin models}, http://arxiv.org/abs/nlin.SI/0410016,
submitted to Journal of Nonlinear Mathematical Physics

\bibitem{KPR} V.B.~Kuznetsov, M.~Petrera, O.~Ragnisco,
{\it Separation of variables and B\"{a}cklund transformations for the symmetric Lagrange top},
Journal of  Physics A {\bf 37} (2004), 8495--8512.

\bibitem{KNS} V.B. Kuznetsov, F.W. Nijhoff and E.K. Sklyanin, {\it 
Separation of variables for the  Ruijsenaars system}, 
Commun. Math. Phys. {\bf 189} (1997), 855--877.

\bibitem{KS5} V.B.~Kuznetsov and E.K.~Sklyanin, {\it On B\"{a}cklund transformations for many-body
systems,} Journal of  Physics A {\bf 31} (1998) 2241--2251.

\bibitem{KV00} V.B.~Kuznetsov and P.~Vanhaecke,
{\it B\"acklund transformations for finite-di\-men\-si\-o\-nal integrable
systems: a geometric approach}, Journal of Geometry and Physics {\bf 44} (2002), 1--40.

\bibitem{RSTS} A.G. Reyman, M.A. Semenov-Tian-Shansky,
{\it Group-Theoretical Methods in the Theory of Finite-Dimensional
Integrable Systems} in Dynamical Systems VII, Springer (1994).

\bibitem{Skl38} E.K.~Sklyanin, {\it Separation of variables. New trends}
Progr.\ Theor.\ Phys.\ Suppl.\ {\bf 118} (1995) 35--60.

\end{thebibliography}
\end{document}